\documentclass[journal=jacsat,manuscript=article, layout=twocolumn]{achemso}
\usepackage{graphicx}
\usepackage{amssymb,amsfonts,amsmath}
\usepackage[pdftex,colorlinks=true,linkcolor=blue,citecolor=blue,urlcolor=blue]{hyperref}
\usepackage[version=4]{mhchem}
\usepackage[dvipsnames]{xcolor}
\usepackage{dcolumn}
\usepackage{bm}
\usepackage[utf8]{inputenc}
\usepackage[T1]{fontenc}
\usepackage{mathptmx}
\usepackage{etoolbox}
\usepackage{tabularx}
\usepackage{xfrac}
\usepackage{doi}

\newcolumntype{Y}{>{\centering\arraybackslash}X}

\setkeys{acs}{doi = true}
\SectionNumbersOn

\newcommand{\ilm}{Institut Lumi\`ere Mati\`ere, UMR CNRS 5306, Univ Lyon, Universit\'e Claude Bernard Lyon 1, F-69622 Villeurbanne, France}
\newcommand{\lmi}{Laboratoire des Multimat\'eriaux et Interfaces, UMR CNRS 5615, Univ. Lyon, Universit\'e Claude Bernard Lyon 1, F-69622 Villeurbanne, France}

\title{Neural network approach for a rapid prediction of metal-supported borophene properties}

\author{Pierre Mignon} 
\author{Abdul-Rahman Allouche} 
\affiliation{\ilm}

\author{Neil Richard Innis} 
\author{Colin Bousige} 
\email{colin.bousige@cnrs.fr}
\affiliation{\lmi}

\begin{document}

\begin{abstract}
We develop a high-dimensional neural network potential (NNP) to describe the structural and energetic properties of borophene deposited on silver. This NNP has the accuracy of DFT calculations while achieving computational speedups of several orders of magnitude, allowing the study of extensive structures that may reveal intriguing moiré patterns or surface corrugations. 
We describe an efficient approach to constructing the training data set using an iterative technique known as the ``adaptive learning approach''.
The developed NNP potential is able to produce, with an excellent agreement, the structure, energy and forces of DFT.  
Finally, the calculated stability of various borophene polymorphs, including those not initially included in the training dataset, shows better stabilization for $\nu\sim0.1$ hole density, and in particular for the allotrope $\alpha$ ($\nu=\sfrac{1}{9}$). 
The stability of borophene on the metal surface is shown to depend on its orientation, implying structural corrugation patterns that can only be observed from long time simulations on extended systems. 
The NNP also demonstrates its ability to simulate vibrational densities of states and produce realistic structures, with simulated STM images closely matching the experimental ones.
\end{abstract}


\section*{TOC Graphic}
\noindent\includegraphics[width=\linewidth]{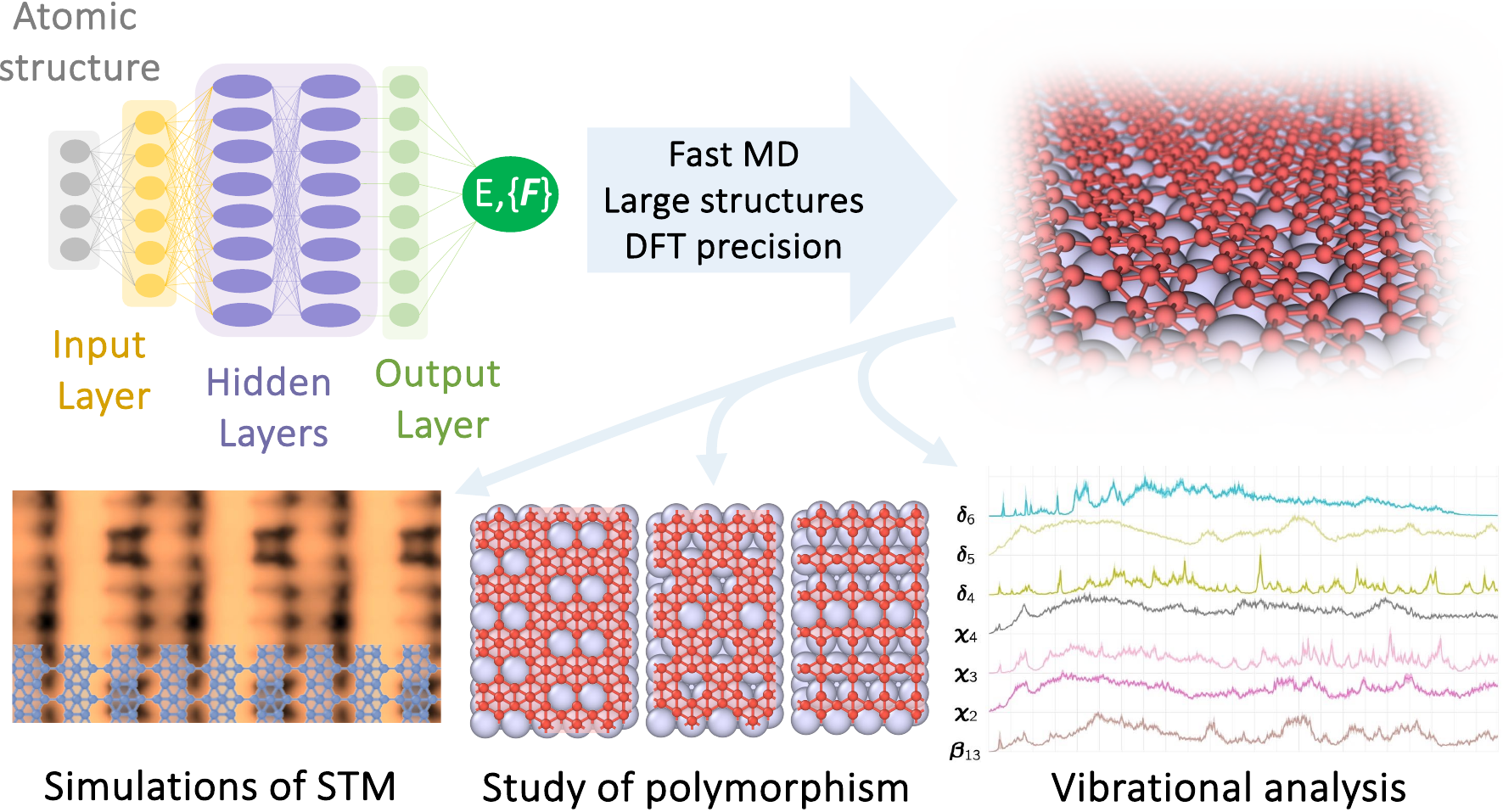}

\section{Introduction}

\begin{figure*}[!ht]
    \centering
    \includegraphics[width=\linewidth]{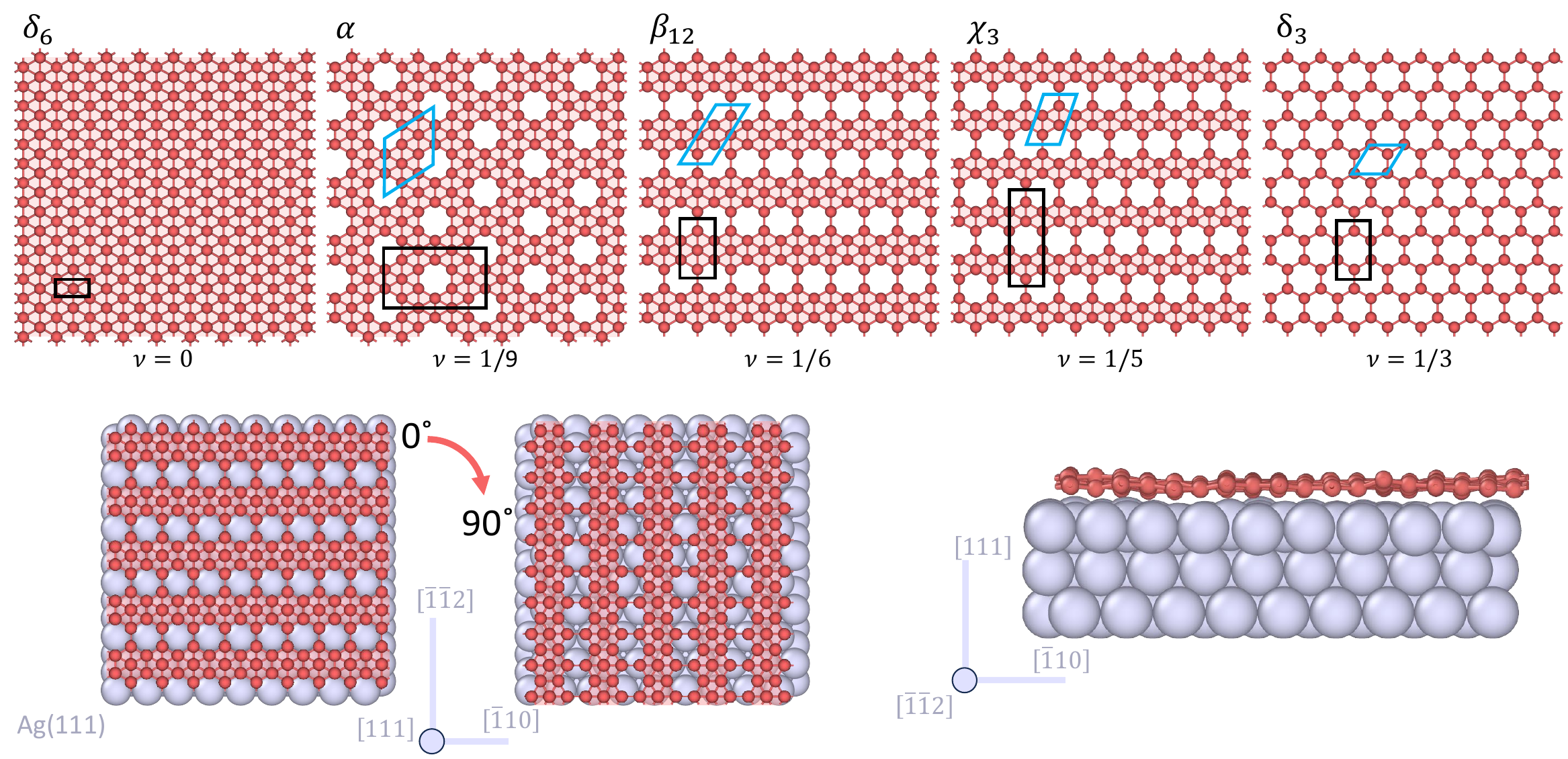}
    \caption{The five different allotropes used in the training dataset, sorted by increasing hole density ($\nu$): $\delta_6$, $\alpha$, $\beta_{12}$, $\chi_3$ and $\delta_3$. 
    All allotropes are initially flat, except for $\delta_6$ which shows a corrugation. The black lines represent the orthogonal unit cells used in the simulations, while the blue ones represent the primitive unit cells. 
    The nomenclature of the different borophene allotropes is based on ref.~[\!\!\citenum{wu_twodimensional_2012}].
    The allotropes are deposited on top of a 3-layer thick Ag(111) or Ag(100) slab, and can be rotated by 90\,$^\circ$ around the $z$ axis while keeping the Ag slab fixed, as shown for example with $\beta_{12}$ on Ag(111).
    }
    \label{fig-allotropes}
\end{figure*}

The recent synthesis of borophene \cite{mannix_synthesis_2015,feng_experimental_2016}, a one-atom-thick 2D crystal of boron with numerous polymorphs \cite{tang_novel_2007,mannix_synthesis_2015,feng_experimental_2016,wu_twodimensional_2012,penev_polymorphism_2012}, has brought forward a missing piece of the 2D materials bestiary: a partially stable metallic 2D material.
Thanks to its interesting properties, borophene may lead to promising applications such as efficient \cite{lherbier_electronic_2016,liu_monolayer_2016}, flexible \cite{mortazavi_anomalous_2017,pang_superstretchable_2016,zhang_elasticity_2017} and transparent \cite{lherbier_electronic_2016} electronics, optoelectronic devices \cite{huang_twodimensional_2017,lian_exotic_2018}, or dense ionic batteries \cite{liu_monolayer_2016,jiang_borophene_2016,li_high_2017,liang_borophene_2017,rao_ultrahigh_2017,zhang_borophene_2017}. 
In addition to its numerous properties\cite{mannix_borophene_2018,hou_borophene_2020,kaneti_borophene_2021,ou_emergence_2021}, borophene shows a high degree of polymorphism, its allotropes being stabilized by the introduction of periodically distributed hexagonal holes into the triangular lattice structure \cite{tang_novel_2007,mannix_synthesis_2015,feng_experimental_2016,wu_twodimensional_2012,penev_polymorphism_2012} (Fig.~\ref{fig-allotropes}).
If there are infinite ways to arrange these hexagonal holes, cluster expansion methods \cite{penev_polymorphism_2012,zhang_twodimensional_2015} have shown several structures (with hole densities in the 10-15\%\ range) with cohesive energies within a few meV/atom of the minimum.
Interestingly, all of the above mentioned properties may be modulated by the degree of anisotropy found in the borophene polymorphs, which contributes to the great richness of this material. 
Thus, one may expect to tune some properties such as plasmon emission, electronic and thermal transport, or mechanical resistance \cite{huang_twodimensional_2017,mannix_borophene_2018} by selectively synthesizing a given polymorph -- note that most polymorphs show metallic behavior \cite{mannix_borophene_2018}.
To date, eleven polymorphs of borophene have been experimentally identified \cite{sutter_largescale_2021,radatovic_macroscopic_2022,omambac_segregationenhanced_2021,cuxart_borophenes_2021,mazaheri_chemical_2021,wu_largearea_2019,vinogradov_singlephase_2019,li_experimental_2018,kiraly_borophene_2019,zhong_synthesis_2017,sheng_raman_2019,zhong_metastable_2017,feng_experimental_2016,mannix_synthesis_2015,wu_micrometrescale_2022,tai_synthesis_2015} and their occurrence has been shown to depend on the experimental synthesis conditions used: temperature, presence of annealing, gas flows, substrate orientation, etc. -- which is very promising for our future ability to selectively synthesize a given polymorph for its desired properties.
Although borophene's allotropes might be identified by Raman spectroscopy \cite{massote_electronic_2016} (which is highly dependent on the structure and electronic state of the studied material), the identification of the synthesized allotrope on metal surfaces is not straightforward, as it is usually done by comparing an experimental scanning tunneling microscopy (STM) image with simulated ones \cite{sutter_largescale_2021,radatovic_macroscopic_2022,omambac_segregationenhanced_2021,cuxart_borophenes_2021,mazaheri_chemical_2021,wu_largearea_2019,vinogradov_singlephase_2019,li_experimental_2018,kiraly_borophene_2019,zhong_synthesis_2017,sheng_raman_2019,zhong_metastable_2017,feng_experimental_2016,mannix_synthesis_2015,wu_micrometrescale_2022,tai_synthesis_2015}.

Overall, the experimental characterization of the structural properties and allotropic configuration of borophene is far from straightforward.
Therefore, theoretical studies are essential to understand and predict its properties, and also to properly characterize the synthesized structures by comparing computational and experimental data.
However, most theoretical studies use Density Functional Theory (DFT) calculations, which are very accurate but also very time consuming and limited in the size of the studied model.
Some theoretical studies \cite{mortazavi_anomalous_2017} on free-standing borophene have been carried out using ReaxFF \cite{vanduin_reaxff_2001}, a classical potential that is rather designed for carbon-based systems and not for boron-substrate interaction -- which is rather important since borophene is always grown on a metal such as silver, gold or copper.
Therefore, in this work we have developed a high-dimensional Neural Network Potential (NNP) \cite{behler_generalized_2007,behler_constructing_2015,singraber_parallel_2019,singraber_librarybased_2019} capable of describing the structural and energetic properties of borophene deposited on silver with the accuracy of DFT calculations while drastically reducing the computational time by several orders of magnitude.
In this work, we focus on silver as it is the most commonly used substrate for borophene synthesis \cite{zhong_synthesis_2017,zhong_metastable_2017,feng_experimental_2016, li_selfassembled_2021, li_synthesis_2021,sheng_raman_2019,li_chemically_2022,mannix_synthesis_2015,liu_probing_2022,liu_borophene_2021}, although our methodology can be easily transferred to other metals.

NNPs are a class of machine learning potentials that have been shown to accurately describe the properties of a wide variety of materials \cite{behler_constructing_2015,singraber_librarybased_2019, rowe_accurate_2020,artrith_highdimensional_2012,kobayashi_machine_2021,kobayashi_machine_2022,jacobsen_onthefly_2018,gastegger_machine_2017}, including boron-containing systems \cite{ghorbanfekr_insights_2020}. It is based on the idea that the potential energy surface (PES) of a system can be approximated by a sum of atomic environment contributions, which in turn can be approximated by a sum of smooth atomic density functions.
The parameters of these smooth functions are then fitted to reproduce the DFT energies and forces of a training set of configurations.
Once trained, the NNP can be used to perform highly accurate  Molecular Dynamics (MD) simulations of extended systems at a fraction of the computational cost of DFT calculations. In the present study, this will allow the description of surface corrugation as a function of the borophene allotrope and surface orientation.

The article is organized as follows.
First, we will present the development of the NNP, with a focus on the iterative construction of the training set through an adaptive learning procedure.
Then, we will discuss the validity of the obtained potential through structural and energetic arguments, on extended models of various allotropes outside of the training set.
Finally, we will show that this NNP can be used to study the stability of polymorphs, to simulate vibrational densities of states (VDOS), and to produce realistic structures whose simulated STM micrographs closely match experimental ones.

\section{Models and methods}

\subsection{Generation of borophene allotropes}

To develop a transferable NNP usable on any borophene allotrope, several structures were generated on Ag substrate.
As shown in Fig.~\ref{fig-allotropes}, the primitive unit cells of borophene allotropes are not necessarily orthogonal or with angles of 60\,$^\circ$ (see Fig.~S1 for a description of the complete set of allotropes). 
To facilitate the accommodation of these structures on an fcc (111) or (100) substrate, an orthogonal unit cell was preferred.
The principle is to generate $N_x\!\times\! N_y$ replica of the initial flat two-atom orthogonal cell (dimension $2.81\times1.62$~\AA$^2$ with boron atoms located at $(0,0)$ and (\sfrac{1}{2},\,\sfrac{1}{2})), and then remove a list of selected atoms from this supercell to obtain the desired allotrope.
The sheet, which may be rotated by 90\,$^\circ$ around the $z$ axis, is then placed on top of an orthogonal Ag(111) or Ag(100) slab replicated along the $x$ and $y$ directions to get the substrate cell parameters as close as possible to those of the borophene sheet.
The equilibrium B-Ag distance given by DFT optimization is of 2.45~\AA, but this distance can be varied during structure generation.
The borophene atomic positions in the surface plane are then multiplied by a correction factor to accommodate the underlying silver surface whose lattice parameters determine the ones of the whole system -- resulting in a slight deformation of the borophene lattice.
A python command-line interface and a graphical user interface have been created to facilitate the generation and visualization of the structures, as well as the generation of VASP (or other formats) input files -- it is freely available\cite{colinbousige_boroml_2023}.

\subsection{High-dimensional neural network potential}

We used the high-dimensional feed-forward neural network potential developed by Behler and Parinello \cite{behler_generalized_2007} and implemented in the \texttt{n2p2} v2.2.0 software \cite{singraber_parallel_2019,singraber_librarybased_2019,singraber_compphysvienna_2021}. 
In this method, the input layer corresponds to the geometric descriptors of the system, treated by hidden layers of a neural networks (usually two) made up of a defined number of neurons each. 
A neural network is defined for each element of the system, resulting in atomic energies and forces (output layers). 
Atomic environments, defined around each atom $i$ by shells of radius $r_c$ (cutoff radius), are then described by a vector of radial and angular symmetry functions, $G_i$, which describe the local environment of each atom in the system in terms of 2- and 3-body densities \cite{behler_atomcentered_2011}.
The ensemble of $G$ functions forms the input layer of the NNP.
In this work we used Gaussian radial functions given as:

\begin{equation}
\label{eq-Gradial}
    G_i^{rad} = \sum_j \text{e}^{-\eta\left(r_{ij} - r_s\right)^2}f_c(r_{ij}),
\end{equation}
as well as narrow angular functions given as:
\begin{equation}
\begin{split}
    \label{eq-Gangular}
    G_i^{ang} = 2^{1-\zeta}\sum_{\substack{j,k\neq i\\j<k}}\left(1+\lambda\cos\theta_{ijk}\right)^\zeta \text{e}^{-\eta\left(r_{ij}^2+r_{ik}^2+r_{jk}^2\right)} \times\\
    f_c(r_{ij})f_c(r_{ik})f_c(r_{jk}),
\end{split}
\end{equation}
with $f_c(r)$ the \texttt{CT\_POLY2} polynomial cutoff function, and where $r_{ij}$ is the distance between atom $i$ and atom $j$, $\theta_{ijk}$ the angle between $\overrightarrow{r_{ij}}$ and $\overrightarrow{r_{ik}}$, and $\lambda$, $\zeta$ and $\eta$ are parameters.
For a full description of the NNP used in \texttt{n2p2}, its symmetry functions and optimization procedures, we refer the reader to refs.~[\!\!\citenum{singraber_parallel_2019,singraber_librarybased_2019,singraber_compphysvienna_2021,behler_constructing_2015,behler_generalized_2007,behler_atomcentered_2011,artrith_highdimensional_2012}].

Here we used a set of 22 radial and 30 angular symmetry functions per element, resulting in an input dimension of 104 for the neural network. 
All parameters of the symmetry functions are provided in the SI along with an example input file.
They have been adapted from those used in ref.~[\!\!\citenum{paleico_global_2020}] to describe copper clusters on a ZnO surface, as they should be well suited to the present similar but simpler system.
Note that in all cases the cutoff radius was set to 6.35~\AA: it is large enough to include all atoms in the first coordination sphere of each atom, but small enough to keep the computational cost reasonable.

Unless otherwise noted, we used a neural network with 2 hidden layers of 20 neurons each.
The softplus and linear activation functions for the hidden and output layers have been used, respectively.
The NNPs were optimized using the multi-stream Kalman filter method \cite{singraber_parallel_2019}, which allows for very fast convergence, and the objective functions included both energies and forces.
The dataset was divided into two subsets for training (90\,\%) and validation (10\,\%).

\subsection{Molecular Dynamics with Neural Network Potentials}

The MD-NNP simulations were performed using the LAMMPS simulation software \cite{plimpton_fast_1995} (version 27May2021) with the \texttt{n2p2} \cite{singraber_compphysvienna_2021} interface implemented in the LAMMPS-NNP package \cite{singraber_librarybased_2019}.
In almost all cases (otherwise noticed), simulations are run with a timestep of 0.1~fs and a Nosé-Hoover thermostat with a relaxation time of 100 timesteps.
The Verlet algorithm \cite{verlet_computer_1967} is used for time integration.
Periodic boundary conditions are applied in all directions.
Note that if the MD simulation encounters a structure outside the range of structures represented in the training dataset, the program will issue an extrapolation warning (EW).
These EWs are to be avoided because they signal that the simulation may be heading with the generation of unrealistic structures -- the NNP is good at interpolation but bad at extrapolation.
In this case, the simulations are usually stopped and the structures raising EW are kept for later inclusion in the training dataset (see details below).

During the testing and renewal phase of the NNP dataset construction (detailed below), dynamics are run for 20~ps with a temperature ramp from 200~K up to 1,000~K in either the NVT or NPT ensembles, and atomic positions are recorded every 20~fs.
The simulations are set to stop when 800 EW have been raised, which corresponds to a maximum of 4 structures having raised an EW per simulation.
For these simulations, all atoms are free to move.

For the vibrational analysis, the system is equilibrated for 10~ps in the NVT ensemble before the production run in the NVE ensemble.
The latter is run for 50~ps, and atomic positions and velocities are recorded every 1~fs.
The vibrational densities of states (VDOS) are then calculated from the square norm of the Fourier transform of the velocities using the \texttt{pdos} function from the \texttt{pwtools} Python package \cite{schmerler_elcorto_2021}.
For these simulations, the bottom two layers of the Ag substrate are fixed to mimic the presence of a substrate.

Sample LAMMPS input files and data handling scripts are freely available on Zenodo.\cite{colinbousige_boroml_2023}

\subsection{Construction of the training dataset}

Building the most representative dataset while avoiding over-representation of given atomic configurations and keeping the computation time (and thus the dataset size) as small as possible is actually the most crucial and difficult part of NNP construction.
For this purpose, we have implemented an iterative construction algorithm based on the adaptive learning procedure\cite{botu_adaptive_2015,gastegger_machine_2017,jacobsen_onthefly_2018,li_dependence_2019,behler_constructing_2015}, which allows to build the training dataset by adding only selected structures while keeping the number of DFT calculations to a minimum. 
In the following, the ``dataset'' refers to the selected structures, associated with their DFT-computed forces and energies, providing the references used for training the NNP. 
The ``stock library'' is a set of available structures that might be integrated in the dataset after computing energy and forces at the DFT level. 
We note here that structures are integrated in the dataset only if their energies are negative and the norms of the force vectors are below 25~eV/\AA.
This filtering is performed each time a new structure is calculated with DFT to ensure that no structure with unrealistic energies or forces is included in the dataset.

\begin{figure*}[!ht]
\centering
  \includegraphics[width=\linewidth]{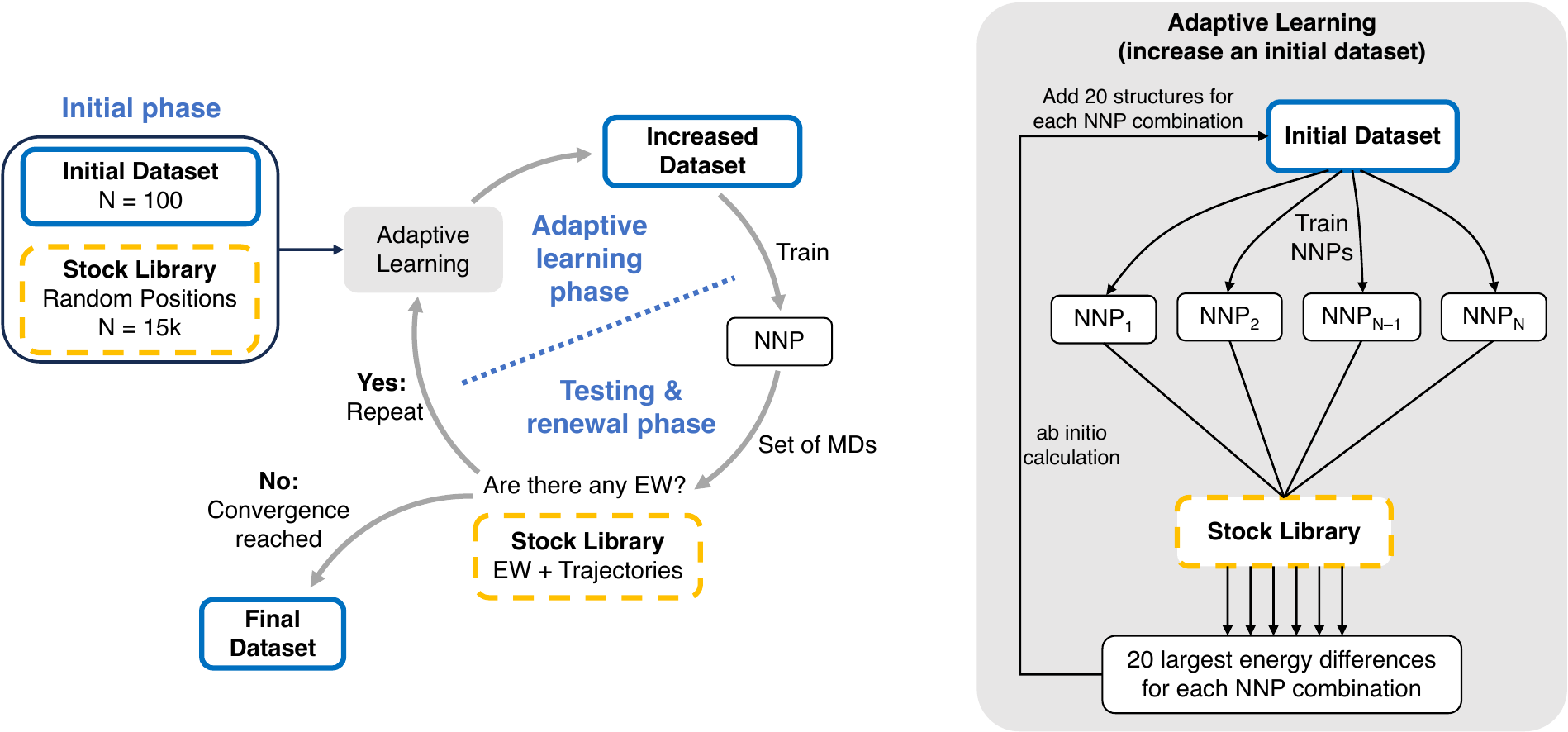}
  \caption{Workflow of the iterative construction algorithm for building the dataset. 
  The procedure is stopped when no extrapolation warnings are found on a set of 50 different test MD simulations on the five training allotropes, going from 200~K to 1,000~K.
  The ``datasets'', framed with a full blue line, are an ensemble of structures associated with their DFT-computed forces and energies.
  The ``stock libraries'', framed with a dashed yellow line, are an ensemble of atomic positions.}
  \label{fig-workflow}
\end{figure*}

The workflow of the iterative construction algorithm is shown in Fig.~\ref{fig-workflow}. 
It consists of an initial phase followed by an alternation of two phases: i) the ``adaptive learning phase'', where the dataset is iteratively enriched by selecting new structures from the stock library, ii) the ``testing and renewal phase'', where the refined NNP is used to perform a series of MD simulations allowing to test its validity and to renew the stock library with ``fresh'' structures. 
Thanks to the adaptive learning procedure, the number of DFT calculations is kept to a minimum and the dataset is enriched with only the most relevant structures.

\paragraph{Initialization}~\\

The initial stock dataset and library are constructed from the 5 allotropes shown in Fig.~\ref{fig-allotropes}. 
These five allotropes were chosen for the training dataset because $\alpha$, $\beta_{12}$, and $\chi_3$ are the most commonly reported allotropes in the experimental literature, and $\delta_{6}$ and $\delta_{3}$ introduce cases where boron atoms are highly or poorly coordinated, respectively. 
They are deposited on Ag(111) and Ag(100) with supercell sizes of $1\!\!\times\!\!1$, $1\!\!\times\!\!2$, $2\!\!\times\!\!1$ and $2\!\!\times\!\!2$ orthogonal unit cells while keeping the number of atoms below 40. 
From these structures, small random atomic displacements of 0.2~\AA\ at most are realized (structural details of these structures are given in Tab.~S1), leading to 100 starting structures that are then computed at the DFT level to create the initial dataset. 
These structures are also used to build an initial stock library of 15k structures by randomly shifting the atoms by a maximum of 0.2~\AA\ and by expanding or compressing the cells by a maximum of 5~\%.

\paragraph{Phase 1: adaptive learning}~\\

The adaptive learning phase then consists in selecting structures from the stock library based on the energy difference calculated from two ghost NNPs with different parameters: two hidden layers each, one NNP with 20 neurons per layer and the other with 15 neurons (see Fig.~\ref{fig-workflow}). 
For a given structure, a large energy difference means that the PES is not well represented from the existing dataset and the structure can potentially be selected and added to the dataset. 
Thus, the energies of all structures in the stock library are computed with these two ghost NNPs, and the 20 structures with the largest energy difference are then selected as new structures to enrich the dataset: DFT energies and forces are thus calculated only for these 20 structures.
The two ghost NNPs are then retrained with this enriched dataset, and the process is repeated until the mean and standard deviation of the energy difference between the two NNPs converge over the entire stock library. 
The training of these ghost NNPs is performed on a small number of epochs (typically 10) to save computational time -- thanks to the Kalman filter method \cite{singraber_parallel_2019}, very fast convergence is achieved anyway. 
The training of the two NNPs, the energy calculations and the DFT computations can be distributed over several nodes and run in parallel on a computing cluster, which makes the whole process quite efficient. 
Since the dataset should be independent of the shape of the NNPs used, we can actually perform this procedure with more than two NNPs (as long as they are well designed) and compare them two-by-two, and convergence is then achieved faster. 
All this procedure is controlled by an in-house python script that is included in the Zenodo archive\cite{colinbousige_boroml_2023}.

\paragraph{Phase 2: testing and renewal of the stock library}~\\

After the adaptive learning phase, we enter the testing and renewal phase. 
We train one of the ghost NNPs above until convergence, and use it to perform 50 test MD simulations on the five training allotropes deposited on the two substrate orientations, in both the NVT and NPT ensembles (see Tab.~S1). 
These MD simulations consist of heating ramps that continuously heat the system from 200~K to 1,000~K in either the NVT or NPT ensemble. 
The goal here is to sample a wide variety of configurations, including high-energy ones, to ensure that the NNP is able to describe the entire PES. 
If too many EW are found, it means that the corresponding atomic configurations are not well represented in the training dataset. 
These simulations are then stopped, and the structures that generated EW (\textit{i.e.} the last 4 in the trajectory) are automatically computed with DFT and included in the new dataset. 
The rest of the trajectories are then concatenated into a new stock library along with their 5\% compression and dilatation analogs, and we enter the adaptive learning phase again.

This alternation of two phases is repeated until no EW is found on the 50 test MD simulations, which in our case happened after 5 iterations when the training dataset reached 9281 structures. 
The initial stock database was increased from the first 15k random structures to $\sim45$k structures, then to $\sim85$k structures, and on to 150k structures in the final step (the maximum without EW).
Note that we could have used the testing and renewal phase to create the initial stock database, but training the NNP for MD simulations on only 100 structures makes no sense.

The goal of the adaptive procedure is to enrich the dataset with new structures describing PES regions that are not yet well represented.
As such, at each new enrichment step, the NNP is able to well reproduce/predict energies and forces for new atomic configurations (\textit{i.e.} interatomic distances and angles). 
Indeed, this is well confirmed by the evolution and broadening of the distribution of atomic configurations (B-B, Ag-Ag and -Ag distances) as the dataset size increases (see Fig.~S2, the smoothing and broadening of the peaks, especially for small distances, which allows to better describe repulsive interactions).
This is clear evidence that the adaptive learning procedure is proceeding with the intended purpose.
We can therefore conclude that the adaptive learning procedure is very efficient, since it allows i) to build the most representative dataset while keeping the computational time (number of DFT calculations) as low as possible, and ii) to define a clear decision threshold for when to stop enriching the dataset.

\paragraph{Final training}~\\

Once the dataset built, the final refined NNP (two hidden layers of 20 neurons each) is trained and convergence  reached after 77 epochs.
The final energy RMSE for training is 26~meV/atom (28~meV/atom for testing), and the final force RMSE for training and testing is 508~meV/\AA\ (see Fig.~S3).
We note here that these values are unusually high for an NNP, but this is due to the fact that we are using a very small dataset with structures that are very different and some are highly energetic because of the induced geometric distortions made by cell compression/expansions on high temperature (until 1,000~K) systems. 
For details see ``distortion'' column of Tab.~S1, the small unit cells used introduce a large amount of strain in the borophene structures (from 3 to 54\%), which is not realistic -- but not a problem per se, as it allows to well describe the limits of repulsive and attractive interactions in the NNP atomic potential.
We were forced to use such small cells to keep the computation time reasonable.
We will see below that applying this NNP to more realistic structures leads to much better RMSEs.
This argument is supported by the fact that much lower MAEs are obtained for energies and forces, since the MAE gives less weight to outliers than the RMSE.
Indeed, we obtain MAEs for energies and forces of 5.4~meV/atom and 203~meV/\AA\ for training, and 7.5~meV/atom and 213~meV/\AA\ for testing -- which are much more reasonable values.
The rather large RMSE observed is thus caused by the occurrence of a few of high-energy-limit structures in the dataset that are fully well described by the NNP.

\subsection{First-principles calculations}

The NNP has been developed on the basis of reference data (energies and forces) from DFT calculations performed with the Vienna Ab initio Simulation Package (VASP) \cite{kresse_initio_1993,kresse_initio_1994,kresse_efficient_1996,kresse_efficiency_1996} using the projector augmented wave (PAW) method to describe ionic cores and valence electrons through a plane wave basis \cite{blochl_projector_1994,kresse_ultrasoft_1999}. 
The Perdew-Burke-Ernzerhof (PBE) form of the generalized gradient approximation (GGA) was used for the exchange and correlation functional\cite{perdew_generalized_1997,perdew_generalized_1996}.
The cutoff energy was fixed at 700~eV.
The bulk Ag unit cell was first optimized using a $9\!\!\times\!\!9\!\!\times\!\!9$ Monkhorts-Pack $\Gamma$-centered mesh to sample the Brillouin zone; this resulted in the cell parameter $a = 4.085$~\AA\ from which the (111) and (100) surface plates were constructed. 
Single point computations were performed on NNP training structures to provide the energy and forces. 
K-point meshes of $7\!\!\times\!\!7\!\!\times\!\!1$ and cutoff energy of 700~eV were chosen from the evaluation of the error computed on energy and forces for different k-point meshes ($3\!\!\times\!\!3\times\!\!1$, $5\!\!\times\!\!5\!\!\times\!\!1$, $7\!\!\times\!\!7\!\!\times\!\!1$, $9\!\!\times\!\!9\!\!\times\!\!1$, $11\!\!\times\!\!11\!\!\times\!\!1$) and cutoff energies (400, 500, 600, 700~eV). 
The selected parameters allowed accurate energy and forces calculations within reasonable computational time (see Figs.~S4-S5). 
All DFT calculations were performed by applying the D3 correction\cite{grimme_D3} to the energy and forces allowing to take into account the van der Waals interactions that are of great importance in the present system. 

The MD simulations used in the validation section are performed on large cells ($\sim25\!\!\times\!\!25\!\!\times\!\!25$~\AA$^3$) containing $\sim500$ atoms.
Thus, the Brillouin zone sampling could be limited to the $\Gamma$ point to keep the computational cost reasonable. 
The energy cutoff is also reduced to the standard value of 400~eV.
For these MD simulations, the thermalization is performed in the NVT ensemble (scaling velocities) for 0.5~ps at 300~K. 
This allows an average temperature of 300~K to be maintained during the 5~ps production run in the NVE ensemble. 
A simulation time step of 1~fs is used. 
The equilibrium of the system is checked and confirmed by verifying that the energy of the system remains stable during the equilibration period. 
Also, the temperature of the system remains stable throughout the production run for all simulations, confirming that the system is well pre-equilibrated. 

\section{Results and discussion}

\subsection{Validation of the model}

The refined NNP is validated by structural and energetic comparison with DFT calculations. 
MD simulations were performed on six borophene allotropes deposited on Ag(111), namely $\alpha$, $\alpha_1$, $\beta_{12}$, $\beta_{13}$, $\chi_2$, and $\chi_3$ (see Fig.~S1 for their structure), using either DFT or the NNP, with the parameters described in the Methods section.
The cell size is set to $\sim25$ \AA\ per side, which is 3 to 5 times larger than those used for the structures in the training dataset, resulting in cells containing $\sim500$ atoms each. 
This allows the stress on the borophene sheets to be reduced with respect to the smaller structures in the training dataset, since a maximum of 3.5\% adjustment of the borophene supercell dimensions on the replicated substrate unit cell has been applied -- the exact cell size for this depends on the allotrope (see Tab.~S2 for all structural details). 
We recall here that only the $\alpha$, $\beta_{12}$ and $\chi_3$ structures are included in the training dataset, with a maximum of 40 atoms per structure (see Tab.~S1). 
In both NNP and DFT cases, the MD is run on a 0.5~ps NVT thermalization at 300~K and 5~ps NVE production (1~fs time step in both cases), and images are saved every 1~fs. 
The initial structure for the MD-NNP is taken as the first one from the MD-DFT production run.
We finally note that each MD-DFT simulation took about 5 days to run on four nodes with 40 cores each, while the MD-NNP simulations ran in less than one hour on one unique node of 40 cores.

\paragraph{Structure}~\\

\begin{figure*}[!ht]
    \centering
    \includegraphics[width=\linewidth]{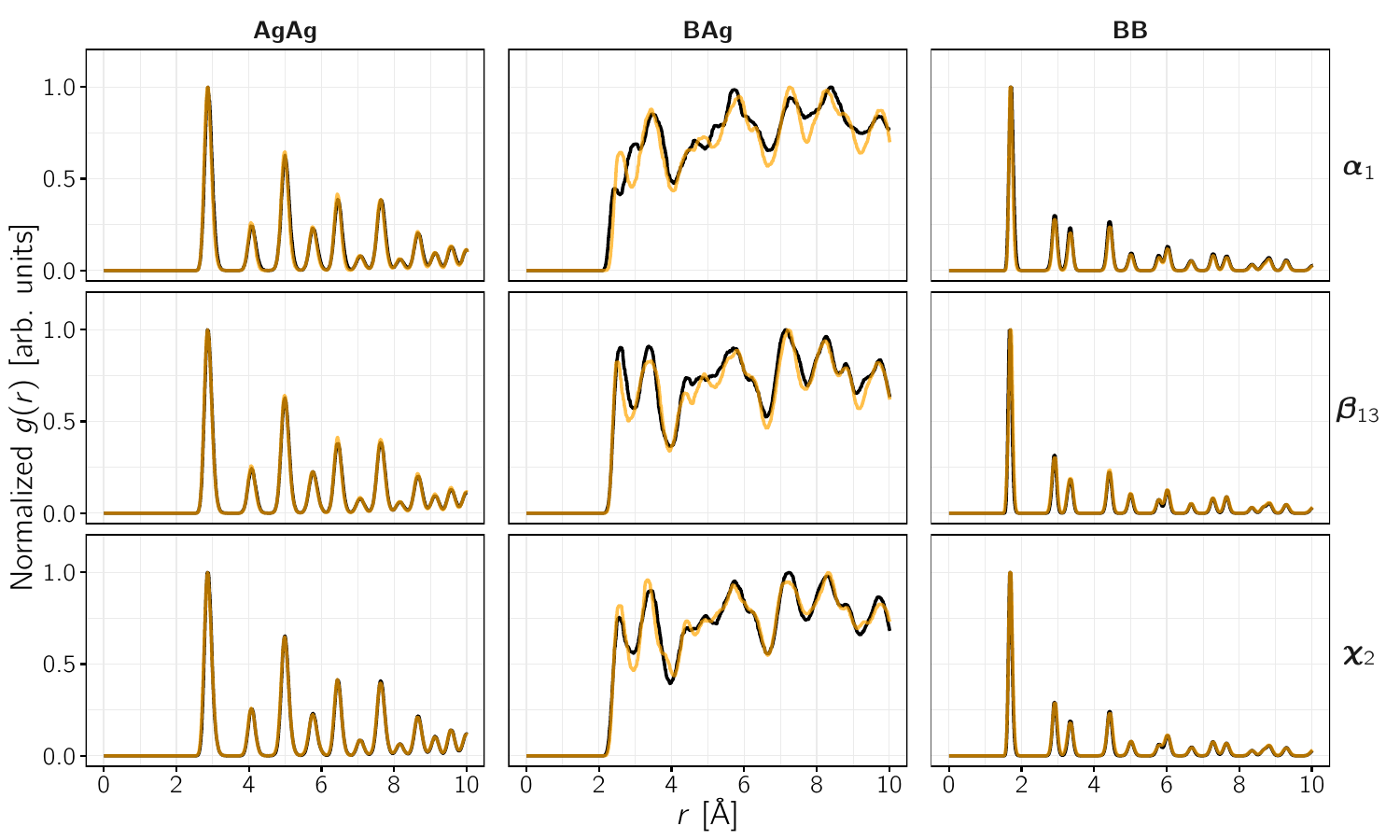}
    \caption{Comparison of partial $g(r)$ calculated on NVE molecular dynamics trajectories at 300~K using DFT (black) or the NNP (orange), for three borophene allotropes on Ag(111): $\alpha_1$, $\beta_{13}$, and $\chi_2$. 
    These allotropes are not included in the NNP training dataset. 
    In all cases, the Ag substrate is composed of three layers, with the bottom two fixed and the lateral cell size $\sim25$~\AA\ (see Tab.~S2 for all structural details).
    }
    \label{fig-GofR}
\end{figure*}

The time-averaged partial radial distribution functions, $g(r)$, for the three allotropes outside the training dataset are shown in Fig.~\ref{fig-GofR} (see Fig.~S6 for all six tested structures) -- note that the Ag-Ag data are filtered to remove the Dirac peaks due to the two lower Ag layers, which are fixed, but otherwise all atoms contribute to the calculation of $g(r)$.

For the Ag-Ag and B-B spatial distributions, Fig.~\ref{fig-GofR} shows that the NNP reproduces the DFT results extremely well up to 10~\AA\ (similar observations are made for allotropes used in the training dataset Fig.~S6).
For the B-Ag case, the agreement is also mostly excellent for all allotropes, the main differences coming from relative peak intensities and widths.
This is probably due to the fact that the initial velocities for the MDs are randomized, resulting in a motion of the borophene sheet on top of the silver slab slightly different in both cases. 
Also, the thermalization method is different in both cases (velocity scaling for MD-DFT, vs. Nosé-Hover for MD-NNP).
It has to be noticed that the boron sheet and Ag surface interact through van der Waals interactions at a distance of about 2.5 ~\AA. 
Compared to most developed NNPs in the literature dealing with covalently bonded materials and better defined PES minima, here the shallow form of the PES due to B-Ag interactions is very well reproduced as seen from the first minima of the BAg radial distribution function. 
Therefore, we can conclude that the NNP is able to reproduce very closely the structures obtained with DFT, despite the fact that the training dataset contains only three of the six tested allotropes and that the tested structures contain $\sim10$ times more atoms.

\paragraph{Energy and forces}~\\

To further validate our NNP, we also compared the energies and forces obtained from the MD-DFT simulations with the MD-NNP ones computed on the same geometries.
Focus is made on the relative variations of the energies, \textit{i.e.} the energy RMSE$^*$ between the DFT and the shifted NNP energy. 
Indeed, the NNP energies, trained from $7\!\!\times\!\!7\!\!\times\!\!1$ k-point grid with a 700~eV cutoff, are naturally shifted compared to the MD-DFT energies, as they are computed on a single k-point and a 400~eV cutoff (see Tab.~S3 for these values, and Fig.~S8 for the time evolution of the energies).

Table~\ref{tab-rmse_test} shows that the energies RMSE$^*$ for all allotropes are about 1~meV/atom, which is excellent for such large structures (we recall that the NNP was trained on structures more than 10 times smaller than these) and for an NNP to which a 6.35~\AA\ cutoff was applied.
Indeed, the NNP energies follow the relative evolution of the DFT energies very closely, in agreement with the small RMSE$^*$ (see Fig.~S8).
This confirms that the NNP is able to accurately reproduce the PES at the level of DFT calculations within the commonly admitted error range of DFT methods, which is the most important aspect.

\begin{table}[!ht]
    \caption{
    Resulting energies and forces RMSE$^*$ for the six test borophene allotropes on Ag(111) along their MD-DFT trajectories. 
    The allotropes in the first group are used in the training dataset, the others are not. 
    The energies RMSE$^*$ are calculated by correcting the NNP energies by the MAE between the NNP and DFT energies (about 10~meV/at).
    }
\label{tab-rmse_test}
\begin{center}
    \begin{tabularx}{\linewidth}{@{}Y|Y|Y@{}}
        Allotrope & Energies RMSE$^*$ [meV/at] & Forces RMSE [meV/\AA] \\
    \hline
    $\alpha$     &  1.22   &  261 \\
    $\beta_{12}$ &  0.807  &  132 \\
    $\chi_3$     &  0.774  &  165 \\
    \hline
    $\alpha_1$   &  1.47   &  299 \\
    $\beta_{13}$ &  1.50   &  337 \\
    $\chi_2$     &  0.929  &  304  \\
\end{tabularx}
\end{center}
\end{table}

Regarding the forces, their RMSE (Tab.~\ref{tab-rmse_test}) are much reduced with respect to the training ones and are in the range of the generally accepted force RMSE for a reliable NNP \cite{paleico_global_2020}. This is due to the fact that the structures encountered along the MD-DFT are all physically sound and less stressed than those present in the training dataset.
The detailed time evolution of the norm of the force vectors for a few atoms along the MD-DFT trajectories computed with DFT and NNP can be found in Fig.~S9.
In addition to the well reproduced shape of the PES from computed energies, this shows that the evolution of any system from MD-NNP simulations allows to explore phase space with an accuracy comparable to DFT ones. 

In conclusion, we have shown that the NNP is able to reproduce the DFT results very accurately in terms of structure, energy and forces, both on the allotropes on which it was trained and on others -- and the training was performed on structures with $\sim10$ times fewer atoms than the ones tested here.
This validates the NNP and allows us to use it to perform MD simulations on large systems with allotropes it was not trained on, which we will do in the next section.

\subsection{Stability analysis}

\begin{figure}[!ht]
    \centering
    \includegraphics[width=\linewidth]{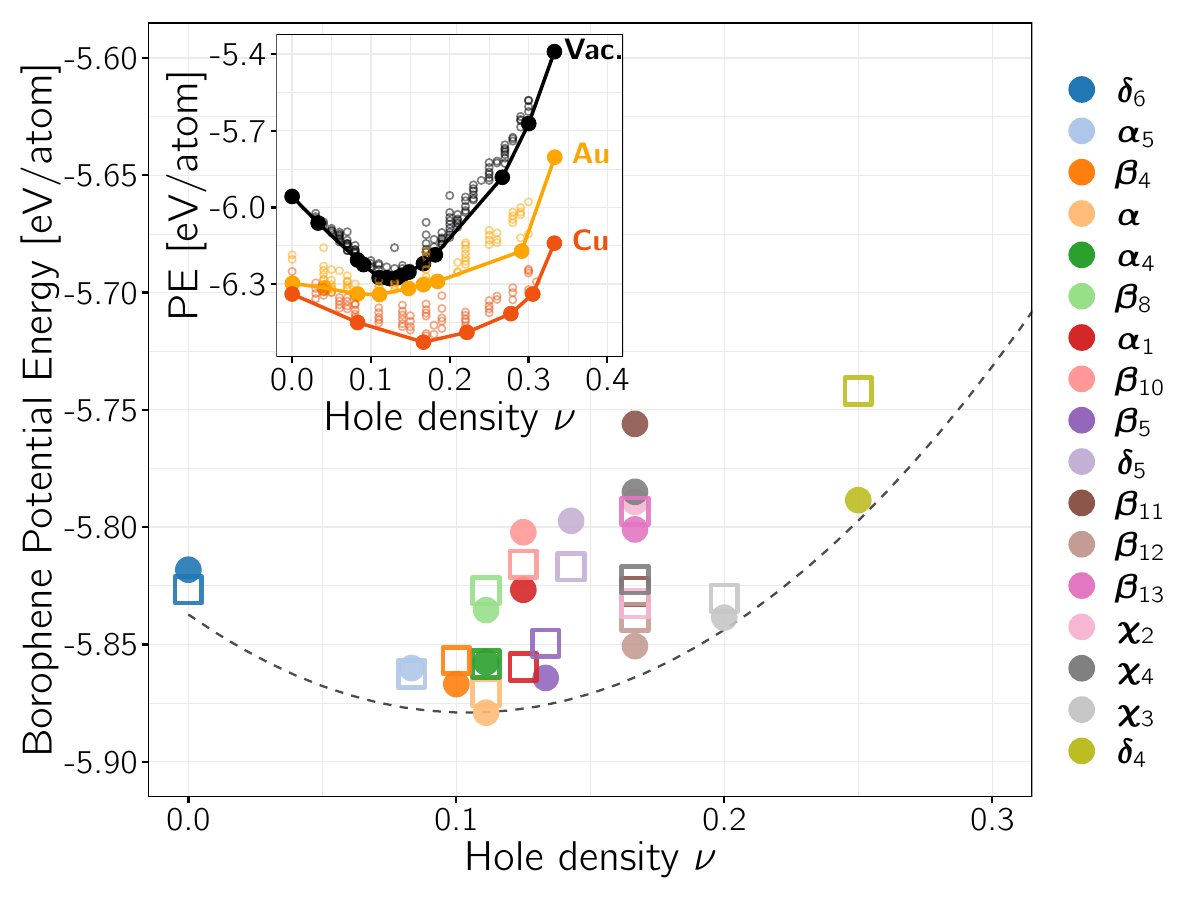}
    \caption{
    Average potential energies of borophene sheets for 17 different stable allotropes on Ag(111) as a function of their hole density $\nu$ after MD relaxation. 
    The full circles are for structures with a 0\,$^\circ$ rotation with respect to the substrate, the empty squares for a 90\,$^\circ$ rotation. 
    All structures have a lattice dimension of at least $25\!\!\times\!\!25\!\!\times\!\!25$~\AA$^3$ and contain about 500 atoms (the lateral size varies from structure to structure in order to keep the borophene distortion below 4\,\% -- details of the structures are given in the SI). 
    The dashed line is a guide to the eye, highlighting the $\left(\nu-\sfrac{1}{9}\right)^2$ trend.
    The inset reproduces data from ref.~[\!\!\citenum{zhang_twodimensional_2015}] and shows the borophene potential energies of free-standing, gold-supported, and copper-supported borophene allotropes as a function of $\nu$. 
    The full signs correspond to DFT calculations, while the empty ones come from the cluster expansion method.
    }
    \label{fig-stability}
\end{figure}

Using the NNP, we performed a stability analysis of 19 different borophene allotropes on Ag(111).
Figure~\ref{fig-stability} shows the average potential energies of the boron atoms for each of these allotropes as a function of their hole density and angular configuration (0\,$^\circ$ or 90\,$^\circ$, as defined in Fig.~\ref{fig-allotropes}).
These energies are averaged over a 5~ps NVE production run after 10~ps thermalization at 300~K: the sheets have thus been allowed to relax on the substrate and buckle out of plane. 
Two of the tested structures are omitted in Fig.~\ref{fig-stability} because of their instability: $\delta_3$ rearranges rapidly during thermalization into a disordered phase with regions resembling $\chi_3$ and others with large holes, and $\alpha_2$ tends to crumple upon itself. 
It has to be noticed that these two allotropes have never been reported on silver.

Very interestingly, Fig.~\ref{fig-stability} shows that the most stable structures are those with $\nu\sim0.1$, and especially the allotrope $\alpha$ ($\nu=\sfrac{1}{9}$). 
In particular the minimum stability profiles (solid/dashed lines) are in very good agreement with that obtained from static DFT calculations and cluster expansion methods\cite{penev_polymorphism_2012,zhang_twodimensional_2015}. Indeed, a minimum is also found for $\nu=\sfrac{1}{9}$ for free-standing or gold-supported borophene -- it shifts to $\nu=\sfrac{1}{6}$ for copper (cf. the inset of Fig.~\ref{fig-stability}). 

It has to be noticed that in our simulations the cell size is much larger and the stability values are averaged over 300~K MD simulations, allowing to describe the corrugation of the borophene sheet above the silver surface. 
This explains the loss of the stability for given allotropes ($\beta_{11}$, $\chi_{4}$, $\beta_{13}$, $\delta_{5}$, $\chi_{2}$) lying rather far above the minimum stability profile. This is thus induced from the dynamic borophene structure deformation that was not taken into account from static DFT calculations. 
Thus, in addition to the fact that our simulations show that the minimum stability for $\nu\sim0.1$ is respected, we observe and describe particular dynamic structural accommodations of given borophene allotropes upon interaction with a metal surface. 

Experimentally, the most commonly reported allotropes on Ag(111) are $\beta_{12}$ ($\nu=\sfrac{1}{6}$) and $\chi_{3}$ ($\nu=\sfrac{1}{5}$), however, it is possible to favor one or the other by playing with annealing times and temperatures\cite{feng_experimental_2016,sheng_raman_2019,li_chemically_2022,mannix_synthesis_2015,zhong_metastable_2017,liu_probing_2022,liu_borophene_2021,li_selfassembled_2021,li_synthesis_2021}, showing that these allotropes are metastable.
We recall that our simulation results are obtained from MD at 300~K, which does not take into account the synthesis pathway, and they are also performed on a limited lateral size, which naturally introduces stress in the borophene lattice.
It would thus be interesting to anneal at larger temperatures and/or over longer periods these allotropes to see whether the $\alpha$ one can be obtained.

We note here that the good agreement between our NNP and the DFT and cluster expansion methods\cite{penev_polymorphism_2012,zhang_twodimensional_2015} is a further confirmation that our NNP is sound and can be used reliably to describe the arrangement of B atoms on the surface regardless of the hole density, as well as to compute the relative energies of different allotropes. 
Moreover, we emphasize that all interatomic interactions are very well represented by the NNP -- which was not a given, considering that B-Ag is a non-bonded, \textit{i.e.} long-range interaction close to the cutoff limit.

From Fig.~\ref{fig-stability}, some allotropes show a large difference in stability upon boron sheet rotation (difference between circle and squares for a given allotrope), this is particularly the case for $\delta_{4}$ and $\beta_{11}$. 
This difference is however not correlated to hole density (see Fig.~S10) neither to the change in borophene sheet distortion due to the rotation (see Tabs.~S4 and S5, $\delta_{4}$ shows almost the lower change). 
Therefore, this shows that for stability evaluations various configurations should always be considered when seeking to identify a given allotrope.
Moreover, it is observed that there logically exists a correlation between hole density and B sheet corrugation over the Ag surface (see Fig.~S11), showing a flatter borophene layer for increasing hole density. 
This is however observed only for the 0\,$^\circ$ configurations for which a positive distortion of the B sheet has been applied for matching the Ag cell dimensions. 
In the case of the 90\,$^\circ$ rotated configurations, the correlation between hole density and surface corrugation is not respected. 
Indeed, for these structures, the borophene sheet is always more corrugated as compared to the 0\,$^\circ$ configurations, which is due to a compressing distortion of the B sheet induced by the matching. 
Nevertheless, taken all together, these data show that the borophene stability above the metallic surface is correlated to the stability of the free borophene allotrope (computed DFT values \cite{penev_polymorphism_2012,zhang_twodimensional_2015}) and to the hole density, but it is also tuned by the geometrical rearrangement of the B sheet on the surface which significantly modulate its stabilization.

\subsection{Vibrational analysis}

\begin{figure}[htp]
    \centering
    \includegraphics[width=\linewidth]{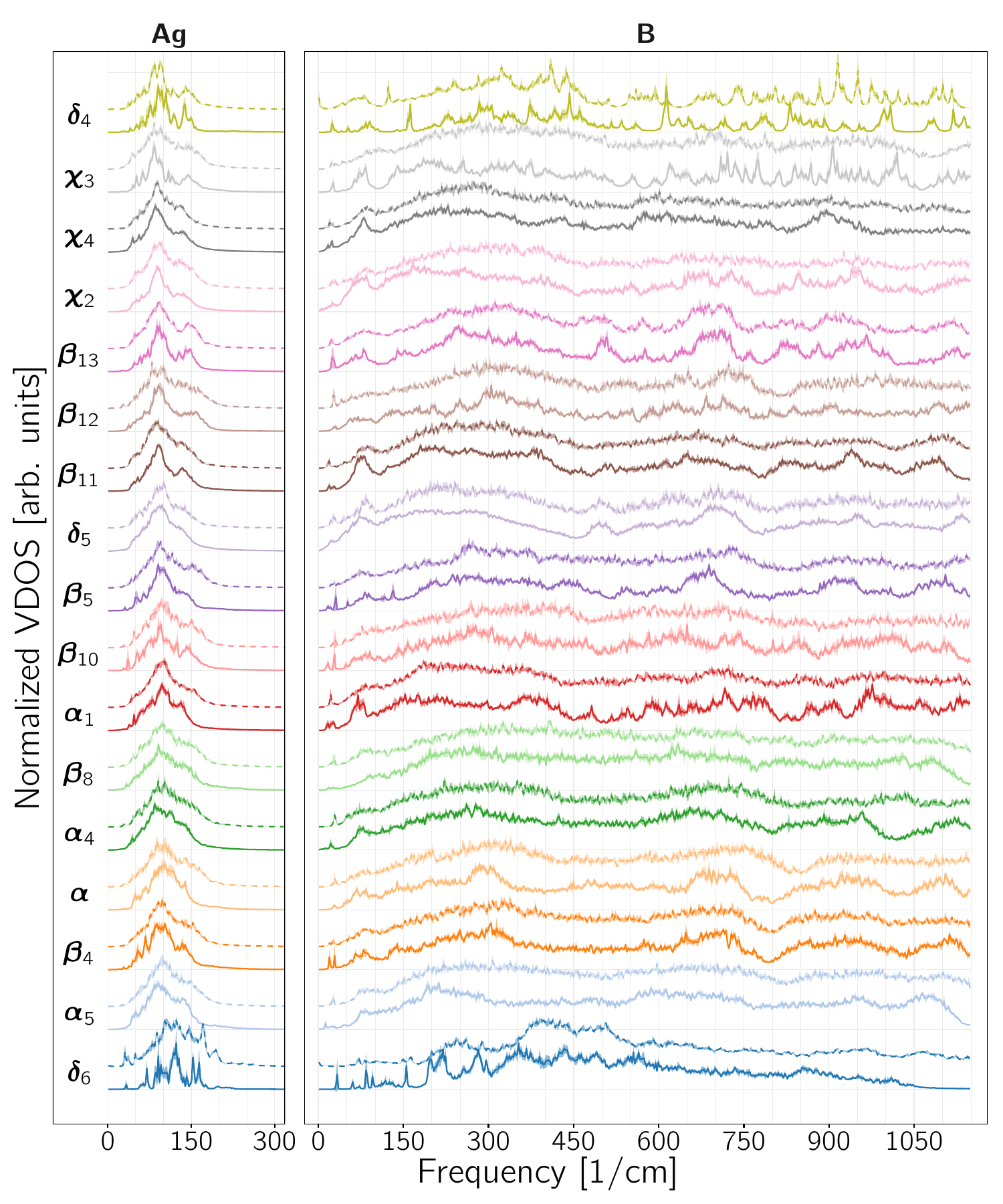}
    \caption{Comparison of the normalized vibrational densities of states (VDOS) for the 17 stable allotropes, as calculated from the silver- or boron-only atomic velocities obtained with an MD-NNP. For each allotrope, the lower and upper curves correspond to the 0\,$^\circ$ and 90\,$^\circ$ configurations, respectively. Details of the structures are given in the SI. The allotropes are ordered by increasing $\nu$ from bottom to top. Thermalization is performed in the NVT ensemble at 300~K, while production is performed in the NVE ensemble.}
    \label{fig-gdos}
\end{figure}

The vibrational density of states (VDOS) of the boron and silver atoms for each allotropes in their 0\,$^\circ$ and 90\,$^\circ$ rotated configurations have been evaluated (Fig.~\ref{fig-gdos}).
50~ps long MD-NNP simulations have been carried out in the NVE ensemble on the 17 stable borophene allotropes on Ag(111) (see Fig.~S1 and Tab.~S4 for structural details), after a 10~ps NVT thermalization at 300~K. 
Again, in all cases, only the Ag atoms in the top layer were allowed to move.

First, the silver VDOS are very similar for all structures, with two peaks at about 100~cm$^{-1}$ and 150~cm$^{-1}$ (with small variations depending on the allotrope), the low energy one being about twice the intensity of the other.
This general shape, independent of the 0\,$^\circ$ or 90\,$^\circ$ configuration, is close to the expected experimental values for bulk silver as measured by inelastic neutron scattering at $\sim125$~cm$^{-1}$ and $\sim180$~cm$^{-1}$ with the same relative intensities \cite{drexel_lattice_1972,dalcorso_densityfunctional_1997}.
This further supports the validity of this approach, and we can assume that the NNP is capable of performing a reliable vibrational analysis on this system.

Figure~\ref{fig-gdos} gathers the VDOS of the boron atoms for the 17 stable allotropes in both 0\,$^\circ$ and 90\,$^\circ$ configurations.
Let us first consider the differences between the allotropes for a single angular configuration, say 0\,$^\circ$. 
We see that most allotropes have very different vibrational profiles with well-defined peaks.
The allotropes that have the broader features are the ones that are the most corrugated (see Fig.~S11 for the $z$ profiles of the different allotropes).
This result is very interesting because it suggests that vibrational analysis could be used to identify the structure of a borophene film on a substrate, since the vibrational profile of the boron atoms should be very different from one allotrope to another.
Now let us look at the differences between the 0\,$^\circ$ and 90\,$^\circ$ configurations.
We can see on Fig.~\ref{fig-gdos} that the VDOS for the 90\,$^\circ$ configuration are generally broader than for the 0\,$^\circ$ one, with less well-defined peaks.
Also, the general shape of the VDOS is often shifted in frequency between the two configurations. 
This frequency shift can be explained by the difference in borophene strain induced by the different borophene distortions in the two configurations (see Tab.~S4 and Tab.~S5).
For the allotropes where the features are broader in the 90\,$^\circ$ configuration, this is probably caused by a more pronounced corrugation in this configuration (see Fig.~S11).
These results show that vibrational analysis can be used to identify borophene allotropes as well as their angular configuration on a substrate, since the vibrational profiles depend on these parameters.

\subsection{STM images of MD-obtained structures}

Using the structures obtained from MD-NNP simulations, it is then possible to compute simulated STM images of the borophene layers on Ag substrate in any configuration from DFT calculations.
Preliminar benchmarking calculations have been carried out in order to check the effect of the number of silver layers as well as the the number of k-points used in the DFT calculation. Results show a low sensibility of the produced STM images and electronic density of the structure with respect to these parameters (Figs.~S13-S14). As such, simulated STM images were obtained from single-point DFT energy calculations at the $\Gamma$ point on a structure containing a single substrate layer in addition to the borophene sheet, in the constant current mode and with a tip placed 2~\AA\ above the top atom.

\begin{figure}[htp]
\centering
    \includegraphics[width=\linewidth]{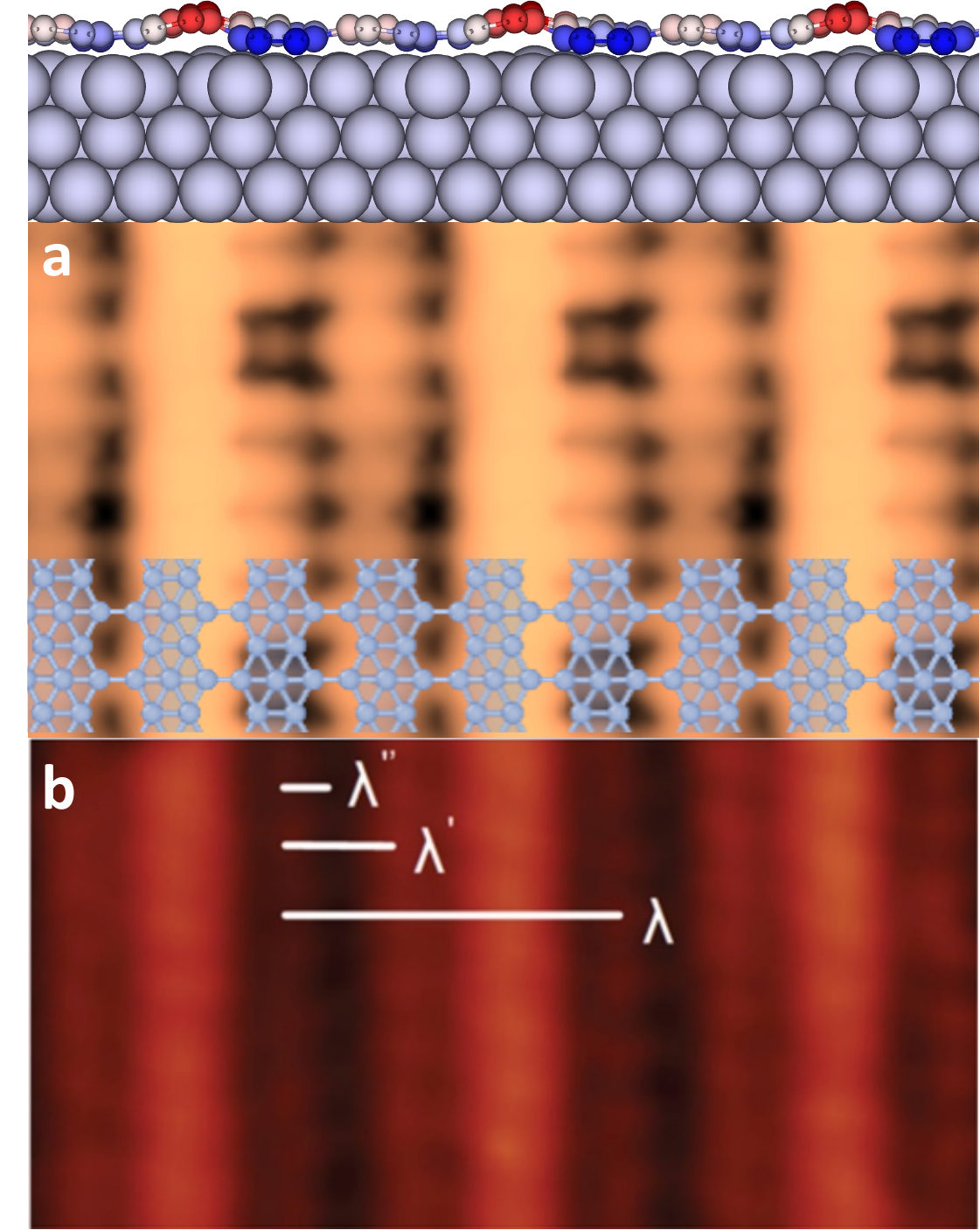}
    \caption{(a) Simulated STM image of the $\beta_{12}$ allotrope on Ag(111) surface in the 90\,$^\circ$ configuration, and (b) experimental STM image of the undulated phase from ref.~[\!\!\citenum{zhang_substrateinduced_2016}] (reproduced with permission. Copyright 2023 American Chemical Society.).
    At the top of panel (a), the boron atoms are colored according to their height, from blue (low) to red (high). This undulated configuration was adopted naturally by the NNP while the two bottom Ag layers were kept fixed.
    } 
    \label{fig-STM}
\end{figure}

Figure~\ref{fig-STM} shows the simulated STM image of the $\beta_{12}$ allotrope on Ag(111) in the 90\,$^\circ$ configuration and compare it with the experimental STM image of the undulated phase obtained experimentally from ref.~[\!\!\citenum{zhang_substrateinduced_2016}].
One can see here a very good agreement between the experimental and simulated STM images, showing that the MD-NNP simulations are able to well reproduce the periodic undulated phase observed experimentally -- without having to introduce a surface deformation of the silver slab.\cite{zhang_substrateinduced_2016}
This suggests that the structures produced by the NNP are very close to the experimental ones, which may be of great help for allotrope identification. 
It has to be noticed that this is made possible because of the large system models considered and allowed from the MD-NNP simulations. 
The use of large lateral sizes is very interesting as it allows the formation of moiré patterns and possibly large wavelength corrugation patterns.
This is very encouraging, as it means that the NNP can be used quite easily to produce large STM images of borophene on metals in any configuration and at any temperature, which is a very useful tool for comparison with experimental images for allotrope identification.

\section{Conclusions}
In this work, we have developed a neural network potential for borophene on silver substrate.
A robust iterative algorithm has been developed to construct the NNP training database, based on the ``adaptive learning'' procedure, which is very general and can be applied to any system. 
The resulting NNP is able to reproduce very accurately the DFT results in terms of structure, energy and forces on large structures, on allotropes that are part of the training set and on others.
This validates the NNP, and allows us to use it to perform long time MD simulations on extended systems with any borophene allotrope, with the accuracy of DFT and for a fraction of its computational cost.
The stability analysis of 19 different borophene allotropes on Ag(111) shows that the most stable structures are those with $\nu\sim0.1$, and in particular the allotrope $\alpha$ ($\nu=\sfrac{1}{9}$). 
We observe that the stability of borophene on the metal surface also depends on its orientation, implying structural corrugation patterns. 
The vibrational analysis of these 19 allotropes shows that the vibrational profiles of the boron atoms are very different from one allotrope to another, and also depend on the angular configuration of the borophene sheet on the substrate.
Finally, we show that the NNP can be used to produce large scale realistic structures of borophene on metals in any configuration and at any temperature, from which large STM images can be simulated, which is a very useful tool for comparison with experimental images for allotrope identification.
In the future, this will be used to build an image database dedicated to the characterization and identification of experimental structures. 
In addition, further work will focus on extending this potential to other metals as well as to multilayer borophene.

\section*{Data Availability Statement}
The data and code that support the findings of this study are openly available in on GitHub and Zenodo\cite{colinbousige_boroml_2023}.

\section*{Declaration of Competing Interest}
The authors declare that they have no known competing financial interests or personal relationships that could have appeared to influence the work reported in this paper.

\section*{Author's contributions}
P.M. contributed to all steps of the study, focusing more on the ab initio calculations.
A.R.A. advised the study concerning the ML parts.
N.R.I. advised the study on borophene chemistry and structure.
C.B. designed and carried out the study.
C.B. and P.M. wrote the manuscript.
All authors discussed and revised the manuscript.

\section*{Acknowledgements}
C.B. acknowledges support by the French National Research Agency grant (ANR-21-CE09-0001-01). 


\providecommand{\latin}[1]{#1}
\makeatletter
\providecommand{\doi}
  {\begingroup\let\do\@makeother\dospecials
  \catcode`\{=1 \catcode`\}=2 \doi@aux}
\providecommand{\doi@aux}[1]{\endgroup\texttt{#1}}
\makeatother
\providecommand*\mcitethebibliography{\thebibliography}
\csname @ifundefined\endcsname{endmcitethebibliography}
  {\let\endmcitethebibliography\endthebibliography}{}

\end{document}